# The "FIP Effect" and the Origins of Solar Energetic Particles and of the Solar Wind

**Donald V. Reames**

Institute for Physical Science and Technology, University of Maryland, College Park, MD 20742-2431 USA, email: dvreames@umd.edu

**Abstract**  We find that the element abundances in solar energetic particles (SEPs) and in the slow solar wind (SSW), relative to those in the photosphere, show different patterns as a function of the first ionization potential (FIP) of the elements. Generally, the SEP and SSW abundances reflect abundance samples of the solar corona, where low-FIP elements, ionized in the chromosphere, are more efficiently conveyed upward to the corona than high-FIP elements that are initially neutral atoms. Abundances of the elements, especially C, P, and S show a crossover from low to high FIP at ≈10 eV in the SEPs but ≈14 eV for the solar wind. Naively this seems to suggest cooler plasma from sunspots beneath active regions. More likely, if the ponderomotive force of Alfvén waves preferentially conveys low-FIP ions into the corona, the source plasma that eventually will be shock-accelerated as SEPs originates in magnetic structures where Alfvén waves resonate with the loop length on closed magnetic field lines. This concentrates FIP fractionation near the top of the chromosphere. Meanwhile, the source of the SSW may lie near the base of diverging open-field lines surrounding, but outside of, active regions, where such resonance does not exist, allowing fractionation throughout the chromosphere. We also find that energetic particles accelerated from the solar wind itself by shock waves at corotating interaction regions (CIRs), generally beyond 1 AU, confirm the FIP pattern of the solar wind.

## 1. Introduction

For many years it has been recognized that the average abundances of the elements in solar energetic particles (SEPs), relative to the corresponding abundances in the solar photosphere, has a characteristic dependence on the first ionization potential (FIP) of the elements (*e.g.* Webber 1977; Meyer 1985). The relative abundances of the elements with FIP < 10 eV (*e.g.* Mg, Si, Fe) are enhanced by a factor of about 4 relative to those with FIP > 10 eV (*e.g.* He, C, O, Ne). This "FIP effect" is understood as an ion-neutral fractionation that occurs as particles expand from the chromosphere up into the corona. The low-FIP elements are easily ionized at photospheric and chromospheric temperatures but those with high FIP are often neutral atoms; the ions are convected upward by the action of Alfvén waves, for example (Laming 2009, 2015), but the neutral atoms are not. All elements become highly ionized on reaching the ≈1 MK corona, but the ionization time for He, at the highest FIP = 24 eV, is the longest.

Meyer recognized that the observed SEP abundances were influenced by two factors. The first was the FIP effect which characterizes the abundances of the corona *before* acceleration and the second was a dependence on the mass-to-charge ratio $A/Q$ of each ion during transport, *after* acceleration, which varied with time and from event to event, as was also clearly shown by Breneman and Stone (1985). The ions in large "gradual" SEP events are accelerated at shock waves driven out from the Sun by coronal mass ejections (CMEs; Kahler *et al.* 1984; Gosling 1993; Cliver, Kahler, and Reames 2004; Lee 2005; Zank, Li, and Verkhoglyadova 2007; Lee, Mewaldt, and Giacalone 2012; Rouillard *et al.* 2011, 2012; Desai and Giacalone 2016; Reames 2017a). The dependence on $A/Q$ may result from rigidity-dependent scattering as the ions spread from the shock (*e.g.* Ng, Reames, and Tylka 2003; Reames 2016a). For example, Fe, with higher $A/Q$ scatters less than O, so Fe/O, at constant velocity will be enhanced early in events and depleted later. Solar rotation can also turn this behavior into a dependence on solar longitude (*e.g.* Reames 2015). Spatial averaging should recover source abundances.

Over the years, our measurement statistics and the sample of SEP events have increased (*e.g.* Reames 1995, 2014) and measurements of the FIP effect in the solar wind have also improved (Bochsler 2009) where a weaker FIP effect is seen in the fast solar wind (FSW), that originates primarily in coronal holes, than in the slow solar wind (SSW), that is often associated with solar active regions. In addition, a property of SEP events, the under-abundance of the element He, has recently become better understood (Reames 2017b) as probable spatial variations in the source plasma. A correction of the He abundance brings the FIP effects of SEPs and the SSW into better agreement. SEP events with He/O ≈ 90 often come from shock acceleration of source plasma with a temperature $T ≈ 3$ MK. The seed population for these events is laced with residual $^3$He-rich, Fe-rich suprathermal ions from previous "impulsive" SEP events in solar active regions (Desai *et al.* 2003; Tylka *et al.* 2005; Reames, Cliver, and Kahler, 2014; Reames 2013, 2016a, 2016b, 2017a, 2017b, 2018). The temperature of 3 MK is actually a property of the residual impulsive suprathermal ions. Events with suppressed He/O involve acceleration of ambient coronal source plasma of < 2 MK (Reames 2017b, 2018).

Do the SEPs and the SSW sample similar regions of the solar corona? How and why do their FIP patterns differ? Are spectroscopic measurements (*e.g.* Schmelz *et al.* 2012; Fludra and Schmelz 1999; Feldman and Widing 2007) of extreme ultraviolet (EUV) and X-ray spectral lines in flares helpful? There are also corotating interaction regions (CIRs) that are formed when FSW streams overtake slow wind, spawning shock acceleration primarily of FSW ions, generally out beyond 1 AU (*e.g.* Richardson 2004). Do abundances of energetic ions from CIRs look like solar wind or like SEPs?

More specifically, one abundance difference that has prevented reconciling SEP and SSW FIP patterns for many years is that of well-measured C/O. Recent abundances are C/O = 0.68 ± 0.07 in both SSW and FSW (Bochsler 2009) and C/O = 0.420 ± 0.010 in SEPs (Reames 2014). None of the 70 individual SEP events studied have C/O > 0.5. Earlier measurements of SEPs and of the solar wind show similar differences, as does the FSW (*e.g.* Gloeckler and Geiss 2007). If C and O are both high-FIP ions, why should their ratio be different from that in the photosphere or from each other? Lack of a convincing answer to this question has stalled our understanding for many years.

## 2. The FIP-Dependence of SEP and SSW Abundances

In Figure 1 we compare the FIP dependence of the ratio of SEP/SSW abundances in the lower panel, while in the upper panel we overlay the usual FIP patterns of the SEP and SSW abundances relative to the photospheric abundances of Caffau *et al.* (2011) and Lodders, Palma, and Gail (2009). The alternative photospheric abundances of Asplund *et al.* (2009) are compared for SEPs by Reames (2015); a choice that has no bearing on our results. In the upper panel we have chosen to multiply the SSW abundances by a factor of 1.2 to improve the relative abundances at high and low FIP, as seen also by the dashed line in the lower panel. These abundances and others used herein are shown in Table 1.

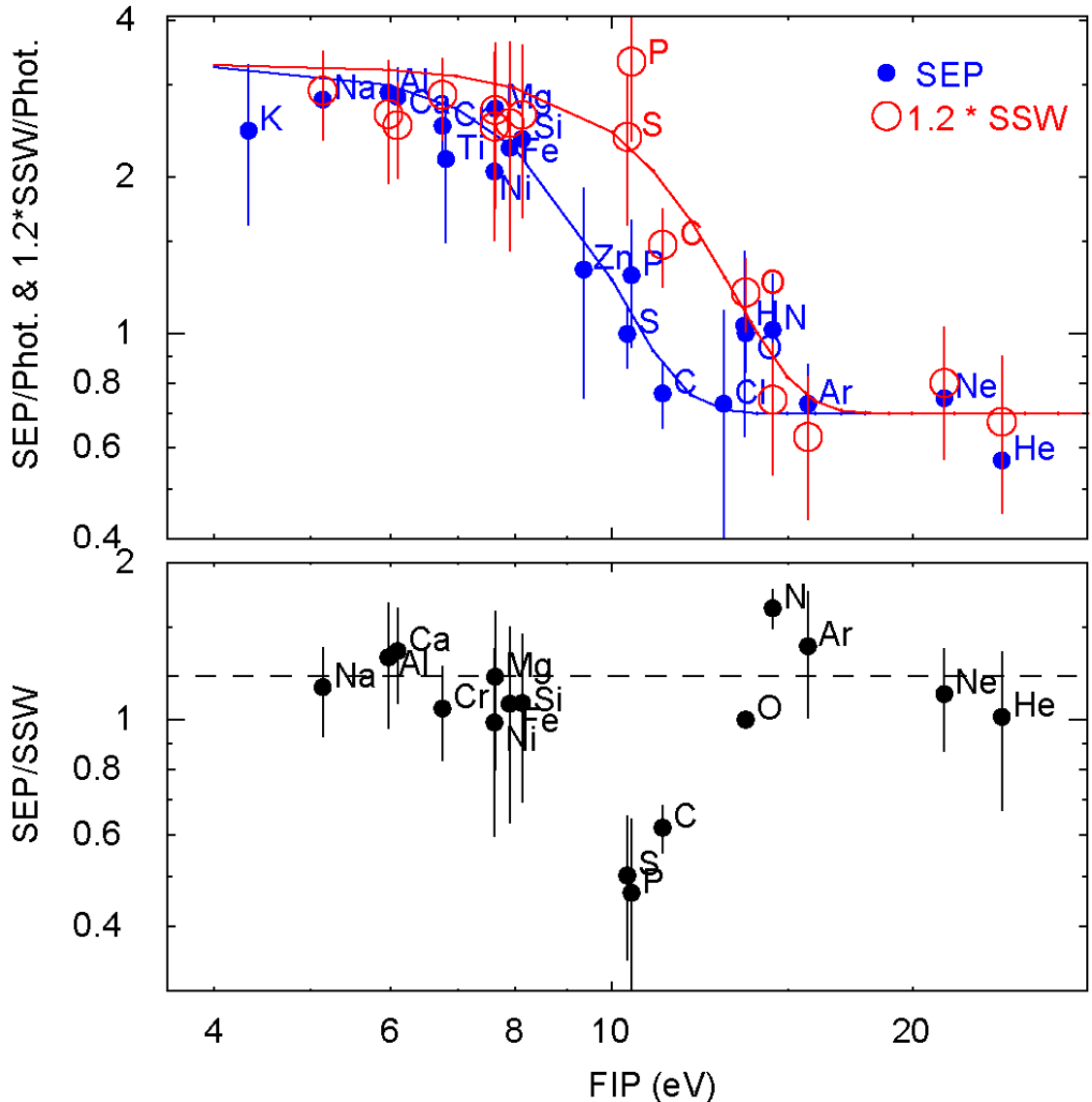

**Figure 1**. The lower panel shows the direct ratio of the element abundances from SEPs to those of the SSW (Bochsler 2009), both normalized at O, as a function of FIP. The dashed line suggests an alternate normalization factor of 1.2. The upper panel shows the SEP/photospheric and 1.2 times the SSW/photospheric abundance ratios as a function of FIP. The curves are empirical and are only used to show the trend of the data.

Table 1 Photospheric, SEP, CIR, SSW, and FSW Abundances.

| | Z | FIP [eV] | Photosphere[1] | SEPs[2] | CIRs[3] | Interstream Solar Wind[4] | Coronal Hole Solar Wind[4] |
|---|---|---|---|---|---|---|---|
| H | 1 | 13.6 | $1.74\times10^{6}$ * | $(\approx1.6\pm0.2)\times10^{6}$ | $(1.81\pm0.24)\times10^{6}$ | – | – |
| He | 2 | 24.6 | $1.6\times10^{5}$ | 91000±5000 | 159000±10000 | 90000±30000 | 75000±20000 |
| C | 6 | 11.3 | 550±76* | 420±10 | 890±36 | 680±70 | 680±70 |
| N | 7 | 14.5 | 126±35* | 128±8 | 140±14 | 78±5 | 114±21 |
| O | 8 | 13.6 | 1000±161* | 1000±10 | 1000±37 | 1000 | 1000 |
| Ne | 10 | 21.6 | 210 | 157±10 | 170±16 | 140±30 | 140±30 |
| Na | 11 | 5.1 | 3.68 | 10.4±1.1 | – | 9.0±1.5 | 5.1±1.4 |
| Mg | 12 | 7.6 | 65.6 | 178±4 | 140±14 | 147±50 | 106±50 |
| Al | 13 | 6.0 | 5.39 | 15.7±1.6 | – | 11.9±3 | 8.1±0.4 |
| Si | 14 | 8.2 | 63.7 | 151±4 | 100±12 | 140±50 | 101±40 |
| P | 15 | 10.5 | 0.529±0.046* | 0.65±0.17 | – | 1.4±0.4 | – |
| S | 16 | 10.4 | 25.1±2.9* | 25±2 | 50±8 | 50±15 | 50±15 |
| Cl | 17 | 13.0 | 0.329 | 0.24±0.1 | – | – | – |
| Ar | 18 | 15.8 | 5.9 | 4.3±0.4 | – | 3.1±0.8 | 3.1±0.4 |
| K | 19 | 4.3 | 0.224±0.046* | 0.55±0.15 | – | – | – |
| Ca | 20 | 6.1 | 3.85 | 11±1 | – | 8.1±1.5 | 5.3±1.0 |
| Ti | 22 | 6.8 | 0.157 | 0.34±0.1 | – | – | – |
| Cr | 24 | 6.8 | 0.834 | 2.1±0.3 | – | 2.0±0.3 | 1.5±0.3 |
| Fe | 26 | 7.9 | 57.6±8.0* | 131±6 | 97±11 | 122±50 | 88±50 |
| Ni | 28 | 7.6 | 3.12 | 6.4±0.6 | – | 6.5±2.5 | – |
| Zn | 30 | 9.4 | 0.083 | 0.11±0.04 | – | – | – |

[1] Lodders, Palme, and Gail (2009)

* Caffau et al. (2011)

[2] Reames (1995, 2014, 2017a)

[3] Reames, Richardson, and Barbier (1991); Reames (1995)

[4] Bochsler (2009)

The remarkable feature of Figure 1 is that the discrepancy between the SEP and SSW abundances seems largely confined to the elements C, P, and S at intermediate values of FIP. It seems that the crossover from low to high FIP occurs at about 14 eV for the SSW and ≈10 eV for the SEPs. Thus the elements C, P, S, and even O (hence the fac-tor of 1.2) behave more like ions in the SSW source but more like neutral atoms in the SEPs.

## 3. The FIP-Dependence of CIR and FSW Abundances

Corotating interaction regions (CIRs) are formed when FSW streams overtake and collide with SSW emitted in that direction earlier in the solar rotation. Two shock waves may be formed, generally beyond 1 AU; the forward shock propagates outward into the SSW and the reverse shock propagates sunward into the FSW. Ions are accelerated mainly from the FSW at the reverse shock which is usually stronger (*e.g.* Richardson *et al.* 1993; Mason and Sanderson 1999; Richardson 2004). Using the data in Table 1, we compare the FIP effect of the CIR and FSW populations in Figure 2.

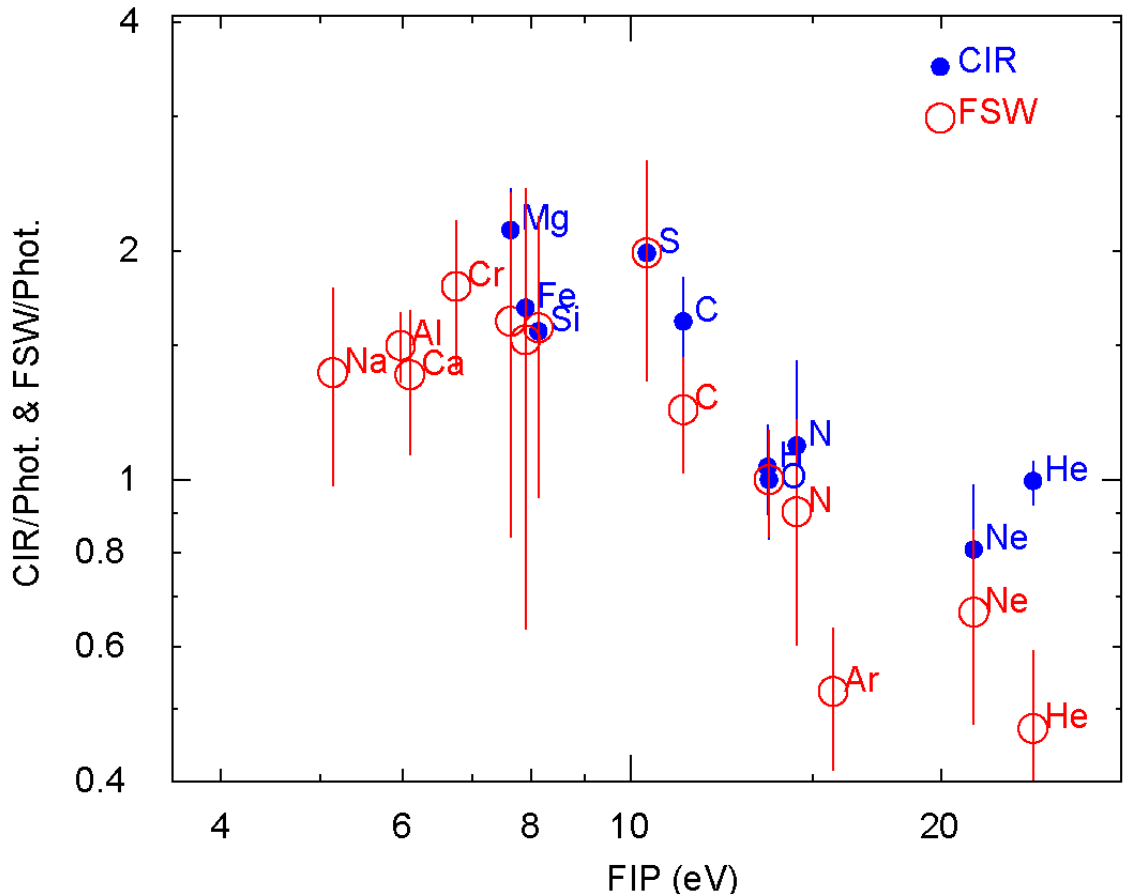

**Figure 2**. The CIR/photospheric and FSW/photospheric abundances, normalized at O, are shown as a function of FIP. C and S behave like the low-FIP ions Mg, Si, and Fe, especially in the CIR population.

With the exception of He, the FIP patterns in Figure 2 are in reasonable agreement. However, variations of He/O are known in the solar wind as functions of time and of solar-wind speed (Collier *et al.* 1996; Bochsler 2007; Rakowsky and Laming 2012). Nevertheless, C and S clearly behave like low-FIP ions, Mg, Si, and Fe, especially in the CIR population, unlike the behavior of the SEPs where C and S behave like N, O, and Ne. There is some evidence that C/O increases with the speed of the high-speed stream (Richardson *et al.* 1993; Mason *et al.* 1997), but this could reflect changes in the seed population that may prefer residual suprathermal SEP ions at the weaker shocks at lower stream speeds.

## 4. Discussion

Brooks, Ugarte and Warren (2016; Abbo *et al.* 2016) have made a full-sun map of FIP bias based upon the S/Si abundance ratio. This map shows a large FIP bias in active regions. Based upon our forgoing analysis, S/Si should certainly show regions of appropriate FIP bias for SEPs, but certainly not for the SSW, since S and Si are both low-FIP elements for the solar wind. Thus S/Si should be unaltered in the source of the SSW. The S/Si map shows the source of the FIP pattern seen by SEPs to be in active regions.

The SEP FIP pattern suggests a source that is cooler, in some sense, than that of the SSW, so that C, S, and P are less likely to be ionized so they behave more like neutral atoms. The model of Laming (2015, 2017) explains the FIP effect in terms of the ponderomotive force of Alfvén waves on ions in the chromosphere and the low corona. This force can differ on closed and open magnetic loops since on closed loops the wavelength can resonate with the loop length. The fractionation is concentrated at the top of the chromosphere where H is becoming ionized if the waves causing the ponderomotive force are in resonance with a coronal loop above (see Figure 8 of Laming 2015). In this case back diffusion of any small neutral fraction restricts fractionation, particularly of C, P, and S which are less ionized than Fe, Mg, and Si. Open field lines are out of resonance and produce ponderomotive force further down where H is neutral, fractionation is easier, and neutral back diffusion less important. Here C, P and S can be fractionated (see Table 4 of Laming 2015). For the SSW, the amplitude of the FIP-bias depends upon the amplitude of slow-mode acoustic waves as shown in Table 4 of Laming (2015).

In Figure 3, the lower panel compares the FIP pattern of the SEPs with the closed field Alfvén-wave model (from Table 3 of Laming 2015) while the upper panel compares the FIP patterns of both the SSW and the CIRs with the open field model (from Table 4 of Laming 2015).

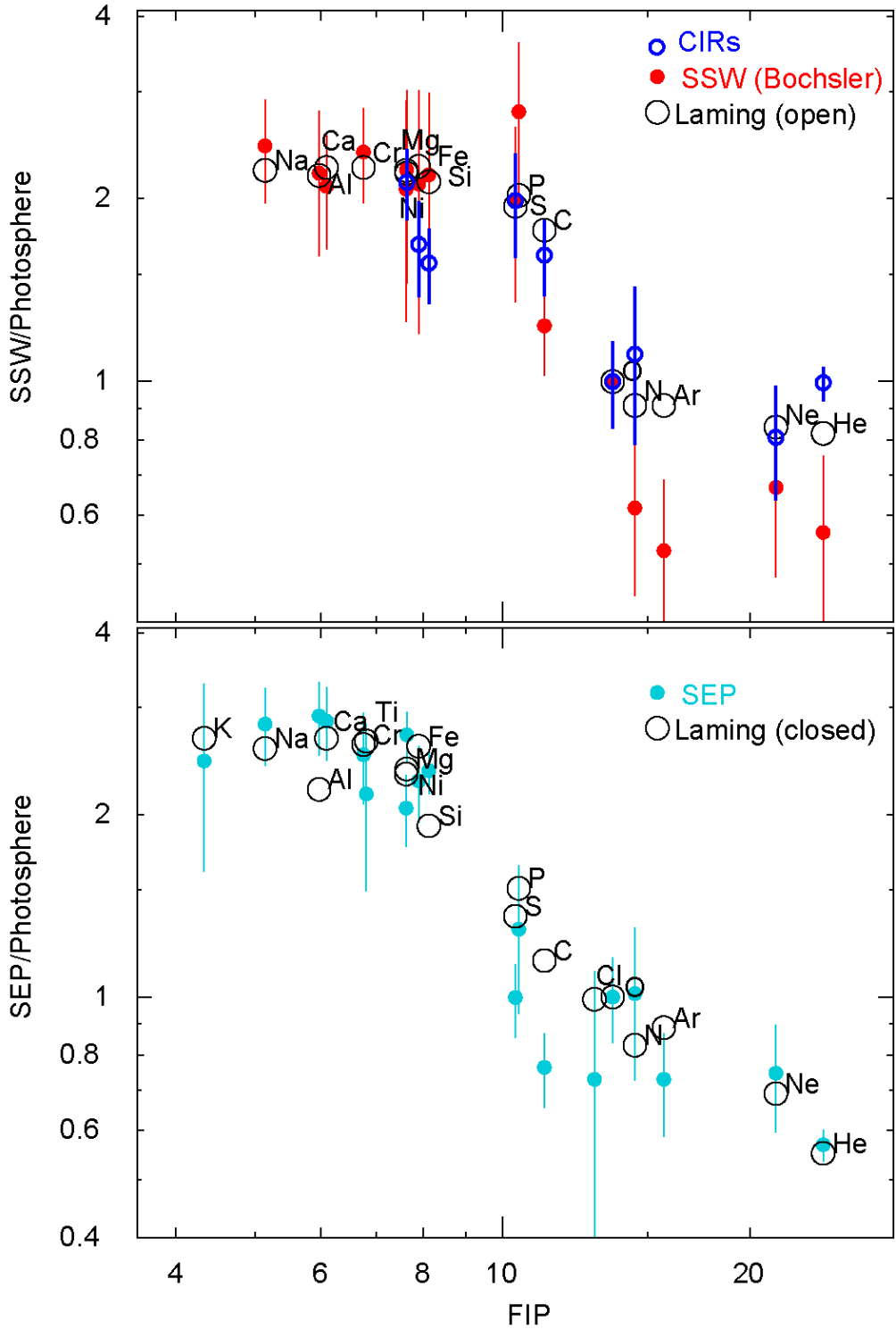

**Figure 3**. The lower panel compares the FIP pattern of SEPs with the closed loop model of Laming (2015, Table 3). The upper panel compares the SSW and CIR FIP patterns the open field model of Laming (2015, Table 4).

The agreement in Figure 3 is generally good, but the theory seems a bit above the SEPs for C and S and below for Si. The SEP abundances, especially C/O, are very well established. Errors for the SSW are larger, but C and the high-FIP elements N, Ar, Ne, and He are consistently below the theory. While the CIR abundances are expected to be more like FSW, rather than SSW, most of the elements fit well, especially for the transition elements, C and P; Si and Fe fall a bit below theory, but Mg agrees well.

X-ray measurements of S, Ca, and Fe in flares seem to show a suppression of S relative to Ca and Fe (Schmeltz *et al.* 2012; Fludra and Schmelz 1999). We should not be surprised that any measurements in flares show the same FIP pattern as SEPs. It is most likely that the suppression of S is also be related to measurements on closed magnetic loops (Laming 2015), but the measurements are surely related to flares and active regions. At a shock wave, ions accelerated from 30 keV amu$^{-1}$ to 3 MeV amu$^{-1}$, for example, have increased their magnetic rigidity and gyroradii by an order of magnitude, so that newly accelerated SEP ions may be able to escape weak trapping on high coronal loops. In addition, the "seed population" for shock acceleration of SEPs is important (see Desai *et al.* 2003; Tylka *et al.* 2005; Laming *et al.* 2013; Reames 2017a). Those SEP events that show higher He/O ratios and 3 MK source plasma temperatures (Reames 2017b) are associated with solar jets from active regions. Other gradual SEP events with source plasma temperatures of 1 – 2 MK (Reames 2016a, 2017a) may involve seed particles from ambient coronal material that was weakly bound on high coronal loops at 2 – 3 solar radii where these SEPs are initially sampled (Reames 2009a, 2009b, 2017a). Many of these are the loops that may be closed for coronal plasma but open for 3 – 10 MeV amu$^{-1}$ SEP ions

SEP events with suppressed values of He/O and source plasma temperatures < 2 MK may involve shock acceleration of plasma from newly formed coronal loops with incomplete He ionization on the fringes of active regions. In any case, SEPs and the SSW must come from different regions of the corona overlying different FIP-dependent processes. Thus, SEPs, at least above a few MeV amu$^{-1}$, are not merely accelerated solar wind; they are a fundamentally different sample of the solar corona. Generally, in the large SEP events, the shock waves begin to sample the corona at 2 – 3 solar radii (Reames 2009a, 2009b; Cliver, Kahler, and Reames 2004) and mostly reaccelerate the same particles farther out.  At lower energies, shock acceleration may continue farther from the Sun and incorporate more of the solar wind plasma, but this is not actually observed (*e.g.* Desai *et al.* 2003).  It has been reported previously (*e.g.* Mewaldt *et al.* 2002), based upon differences in FIP, that SEPs can not be only accelerated solar wind.  However, the present article is the first to characterize the FIP patterns as differences in the location of the crossover between high and low FIP, to discuss the relationship with CIR abundances, and to consider the theoretical connection to open and closed magnetic loops.

Why does C/O differ in SEPs and the SSW?  C behaves as a high-FIP neutral atom in the closed loops in active regions that supply seed particles for SEPs.  However, C is a transition element, partially ionized and partially enhanced during transit to the corona that contributes to the SSW.

SEPs are one of the most complete samples of coronal abundances that we have.  Studying them may provide insight on the origin of these and of other coronal samples as well.

**Acknowledgments:** The author thanks Martin Laming for helpful discussions related to the theory included in this manuscript.

# Disclosure of Potential Conflicts of Interest

The authors declare they have no conflicts of interest.

# References

Abbo, L., Ofman, L., Antiochos, S.K., Hansteen, V.H., Harra, L., Ko, Y.-K. et al.: 2016, Slow solar wind: Observations and modeling *Space Sci. Rev.* **201** 55, doi: 10.1007/s11214-016-0264-1

Asplund, M., Grevesse, N., Sauval, A.J., Scott, P.: 2009, The chemical composition of the sun. *Ann. Rev. Astron. Astrophys*. **47**, 481 doi: 10.1146/annurev.astro.46.060407.145222

Bochsler, P.: 2007, Solar abundances of oxygen and neon derived from solar wind observations, *Astron. Astrophys.* **471** 315, doi: 10.1051/0004-6361:20077772

Bochsler, P.: 2009, Composition of matter in the heliosphere, *Proc. IAU Sympos* **257**, 17, doi:10.1017/S1743921309029044.

Breneman, H.H., Stone, E.C.: 1985, Solar coronal and photospheric abundances from solar energetic particle measurements, *Astrophys. J. Lett.* **299**, L57, doi: 10.1086/184580

Brooks, D.H., Ugarte-Urra, I., Warren, H.P.:2016, Full-Sun observations for identifying the source of the slow solar wind *Nature Comms.* **6**, 5947, doi: 10.1038/ncomms6947

Caffau, E., Ludwig, H.-G., Steffen, M., Freytag, B., Bonofacio, P.: 2011, Solar chemical abundances determined with a CO5BOLD 3D model atmosphere, *Solar Phys.* **268**, 255. doi:10.1007/s11207-010-9541-4

Cliver, E.W., Kahler, S.W., Reames, D.V.: 2004, Coronal shocks and solar energetic proton events, *Astrophys. J.* **605**, 902, doi: 10.1086/382651

Collier, M.R., Hamilton, D.C., Gloeckler, G., Bochsler, P., Sheldon, R.B.: 1996, Neon-20, oxygen-16, and helium-4 densities, temperatures, and suprathermal tails in the solar wind determined with WIND/MASS, *Geophys. Res. Lett.,* **23**, 1191 doi: 10.1029/96GL00621

Desai, M.I., Giacalone, J.: 2016, Large gradual solar energetic particle events, *Living Reviews of Solar Physics*, doi: 10.1007/s41116-016-0002-5.

Desai, M.I., Mason, G.M., Dwyer, J.R., Mazur, J.E., Gold, R.E., Krimigis, S.M., Smith, C.W., Skoug, R.M.: 2003, Evidence for a suprathermal seed population of heavy ions accelerated by interplanetary shocks near 1 AU, *Astrophys. J.* **588**, 1149, doi: 10.1086/374310

Feldman, U., Widing, K.G.: 2007, Spectroscopic measurement of coronal compositions, *Space Sci. Rev.* **130** 115  doi: 10.1007/s11214-007-9157-7

Fludra, A., Schmelz, J. T.: 1999, The absolute coronal abundances of sulfur, calcium, and iron from Yohkoh-BCS flare spectra, *Astron. Astrophys.* **348**, 286.

Gloeckler,G., Geiss, J.: 2007, The composition of the solar wind in polar coronal holes, *Space Sci. Rev.* **130** 139 doi: 10.1007/s11214-007-9189-z

Gosling, J.T.: 1993 The solar flare myth, *J. Geophys. Res.* **98**, 18937 doi: 10.1029/93JA01896

Kahler, S.W., Sheeley, N.R.,Jr., Howard, R.A., Koomen, M.J., Michels, D.J., McGuire R.E., von Rosenvinge, T.T., Reames, D.V.: 1984, Associations between coronal mass ejections and solar energetic proton events, *J. Geophys. Res.* **89**, 9683, doi: 10.1029/JA089iA11p09683

Laming, J.M.: 2009, Non-WKB models of the first ionization potential effect: implications for solar coronal heating and the coronal helium and neon abundances, *Astrophys. J.* **695**, 954 doi: 10.1088/0004-637X/695/2/954

Laming, J.M.: 2015, The FIP and inverse FIP effects in solar and stellar coronae, *Living Reviews in Solar Physics*, **12**, 2 doi: 10.1007/lrsp-2015-2

Laming, J.M.: 2017, The First Ionization Potential Effect from the Ponderomotive Force: On the Polarization and Coronal Origin of Alfvén Waves *Astrophys J. Lett.* **844** L153 doi: 10.3847/1538-4357/aa7cf1Lee, M.A.: 2005, Coupled hydromagnetic wave excitation and ion acceleration at an evolving coronal/interplanetary shock, *Astrophys. J. Suppl.,* **158**, 38, doi: 10.1086/428753

Laming, J.M. Moses, J.D., Ko, Y.-K. Ng, C.K., Rakowski, C.E.;,Tylka, A.J.: 2013, On the remote detection of suprathermal ions in the solar corona and their role as seeds for solar energetic particle production, *Astrophys. J.* **770** 73 doi: 10.1088/0004-637X/770/1/73

Lee, M.A., Mewaldt, R.A., Giacalone, J.: 2012, Shock acceleration of ions in the heliosphere, *Space Sci. Rev.,* **173**, 247, doi: 10.1007/s11214-012-9932-y

Lodders, K., Palme, H., Gail, H.-P.: 2009, Abundances of the elements in the solar system, In: Trümper, J.E. (ed.) *Landolt-Börnstein, New Series VI/4B*, Springer, Berlin. Chap. **4.4**, 560.

Mason, G.M., Mazur, J.E., Dwyer, J.R., Reames, D.V., von Rosenvinge, T.T.: 1997, New spectral and abundance features of interplanetary heavy ions in corotating interaction regions, *Astrophys, J.* **486** 149 doi: 10.1086/310845

Mason, G.M., Sanderson, T.R.: 1999, CIR associated energetic particles in the inner and middle heliosphere, *Space Sci. Rev.,* **89**, 77 doi: 10.1023/A:1005216516443

Mewaldt, R.A., Cohen, C.M.S., Leske, R.A., Christian, E.R., Cummings, A.C., Stone, E.C., von Rosenvinge, T.T., Wiedenbeck, M.E.: 2002, Fractionation of solar energetic particles and solar wind according to first ionization potential, *Advan. Space Res.* **30** 79 doi: 10.1016/S0273-1177(02)00263-6

Meyer, J.-P.: 1985, The baseline composition of solar energetic particles, *Astrophys. J. Suppl.* **57**, 151, doi: 10.1086/191000

Ng, C.K., Reames, D.V., Tylka, A.J.: 2003, Modeling shock-accelerated solar energetic particles coupled to interplanetary Alfvén waves, *Astrophys. J.* **591**, 461, doi: 10.1086/375293

Rakowsky, C.E., Laming, J. M.: 2012, On the origin of the slow speed solar wind: helium abundance variations *Astrophys. J.* **754**, 65, doi: 10.1088/0004-637X/754/1/65

Reames, D.V.: 1995, Coronal Abundances determined from energetic particles, *Adv. Space Res.* **15** (7), 41.

Reames, D.V.: 2009a, Solar release times of energetic particles in ground-level events, *Astrophys. J.* **693**, 812, doi: 10.1088/0004-637X/693/1/812

Reames, D.V.: 2009b, Solar energetic-particle release times in historic ground-level events, *Astrophys. J.* **706**, 844, doi; 10.1088/0004-637X/706/1/844

Reames, D.V.: 2013, The two sources of solar energetic particles, *Space Sci. Rev.* **175**, 53, doi: 10.1007/s11214-013-9958-9

Reames, D.V.:2014, Element abundances in solar energetic particles and the solar corona, *Solar Phys.*, **289**, 977 doi: 10.1007/s11207-013-0350-4

Reames, D.V.: 2015, What are the sources of solar energetic particles? Element abundances and source plasma temperatures, *Space Sci. Rev.,* **194**: 303, doi: 10.1007/s11214-015-0210-7.

Reames, D.V.: 2016a, Temperature of the source plasma in gradual solar energetic particle events, *Solar Phys.*, **291** 911, doi: 10.1007/s11207-016-0854-9

Reames, D.V.:2016b, The origin of element abundance variations in solar energetic particles, *Solar Phys*, **291** 2099, doi: 10.1007/s11207-016-0942-x,

Reames D.V.: 2017a, *Solar Energetic Particles*, Lecture Notes in Physics **932**, Springer, Berlin, ISBN 978-3-319-50870-2, DOI 10.1007/978-3-319-50871-9.

Reames, D.V., 2017b The abundance of helium in the source plasma of solar energetic particles, *Solar Phys.* **292** 156 doi: 10.1007/s11207-017-1173-5 (arXiv: 1708.05034)

Reames, D.V.: 2018, Abundances, ionization states, temperatures, and FIP in solar energetic particles, *Space Sci. Rev.* submitted, (arxiv: 1709.00741)

Reames, D.V., Cliver, E.W., Kahler, S.W.: 2014, Abundance enhancements in impulsive solar energetic-particle events with associated coronal mass ejections, *Solar Phys*. **289**, 3817 doi: 10.1007/s11207-014-0547-1

Reames, D.V., Richardson, I.G., Barbier,, L.M.:1991, On the differences in element abundances of energetic ions from corotating events and from large solar events, *Astrophys. J. Lett*. **382**, L43 doi: 10.1086/186209

Richardson, I. G.: 2004, Energetic particles and corotating interaction regions in the solar wind, *Space Sci. Rev.* **111** 267 doi: 10.1023/B:SPAC.0000032689.52830.3e

Richardson, I.G., Barbier, L.M., Reames, D.V., von Rosenvinge, T.T.: 1993, Corotating MeV/amu ion enhancements at ≤ 1 AU from 1978 to 1986, *J. Geophys. Res.* **98** 13 doi: 10.1029/92JA01837

Rouillard, A.C., Odstrčil, D., Sheeley, N.R. Jr., Tylka, A.J., Vourlidas, A., Mason, G., Wu, C.-C., Savani, N.P., Wood, B.E., Ng, C.K., et al.: 2011, Interpreting the properties of solar energetic particle events by using combined imaging and modeling of interplanetary shocks, *Astrophys. J.* **735**, 7, doi: 10.1088/0004-637X/735/1/7

Rouillard, A., Sheeley, N.R.Jr., Tylka, A., Vourlidas, A., Ng, C.K., Rakowski, C., Cohen, C.M.S., Mewaldt, R.A., Mason, G.M., Reames, D., et al.: 2012, The longitudinal properties of a solar energetic particle event investigated using modern solar imaging, *Astrophys. J.* **752** 44, doi: 10.1088/0004-637X/752/1/44

Schmelz , J. T., Reames, D. V., von Steiger, R., Basu, S.:2012, Composition of the solar corona, solar wind, and solar energetic particles, Astrophys. J. **755** 33

Tylka, A.J., Cohen, C.M.S., Dietrich, W.F., Lee, M.A., Maclennan, C.G., Mewaldt, R.A., Ng, C.K., Reames, D.V.: 2005, Shock geometry, seed populations, and the origin of variable elemental composition at high energies in large gradual solar particle events, *Astrophys. J.* **625**, 474, doi: 10.1086/429384

Webber, W. R.: 1975, Solar and galactic cosmic ray abundances - A comparison and some comments. *Proc. 14th Int. Cos. Ray Conf*, (Munich), **5**, 1597.

Zank, G.P., Li, G., Verkhoglyadova, O., Particle Acceleration at Interplanetary Shocks*, Space Sci. Rev.* **130**, 255 (2007) doi: 10.1007/s11214-007-9214-2